\documentclass[a4paper,11pt]{article}

\usepackage{jheppub} 
\usepackage[normalem]{ulem}  

\usepackage{feynmp-auto}

\usepackage{float}

\renewcommand{\d}{\textrm{d}}

\newcommand{\ba}{\begin{equation}\begin{aligned}}
\newcommand{\ea}{\end{aligned}\end{equation}}

\renewcommand{\d}{\textrm{d}}

\def\integers          {{\mathbb Z}}

\newcommand{\de}{\partial}

\renewcommand{\d}{\mathrm{d}}

\title{Duality and Axionic Weak Gravity}

\author[a,b]{Stefano Andriolo,}
\author[c]{Tzu-Chen Huang,}
\author[d]{Toshifumi Noumi,}
\author[c,e]{Hirosi Ooguri,}
\author[f]{Gary Shiu}

\affiliation[a]{Department of Physics and Jockey Club Institute for Advanced Study, Hong Kong University of Science and Technology, Hong Kong}
\affiliation[b]{Institute of Theoretical Physics, KU Leuven,
Celestijnenlaan 200D B-3001 Leuven, Belgium}
\affiliation[c]{Walter Burke Institute for Theoretical Physics, California Institute of Technology, Pasadena, CA91125, USA}
\affiliation[d]{Department of Physics, Kobe University, Kobe 657-8501, Japan}
\affiliation[e]{Kavli Institute for the Physics and Mathematics of the Universe (WPI), University of Tokyo, Kashiwa, 277-8583, Japan}
\affiliation[f]{Department of Physics, University of Wisconsin-Madison, Madison, WI 53706, USA}

\emailAdd{stefano.andriolo@kuleuven.be}
\emailAdd{jimmy@caltech.edu}
\emailAdd{tnoumi@phys.sci.kobe-u.ac.jp}
\emailAdd{ooguri@caltech.edu}
\emailAdd{shiu@physics.wisc.edu}

\preprint{CALT-TH 2020-007, IPMU20-0035, KOBE-COSMO-20-04, MAD-TH-20-02}

\abstract{
The axionic weak gravity conjecture predicts the existence of instantons whose actions are less than their charges in appropriate units. We show that the conjecture is satisfied for the axion-dilaton-gravity system if we assume 
duality constraints on the higher derivative corrections in addition to positivity bounds  which follow from unitarity, analyticity, and locality of UV scattering amplitudes. On the other hand, the conjecture does not follow if we assume the positivity bounds only. This presents an example where derivation of the weak gravity conjecture requires 
more detailed UV information than the consistency of scattering amplitudes.
}

\begin{document} 
\setcounter{tocdepth}{2}
\maketitle
\flushbottom

\section{Introduction}

The Swampland program is based on the premise that there are conditions on a low energy gravity theory that
are necessary in order for it to be an effective theory of a consistent quantum gravity with ultra-violet (UV) completion such as string theory 
but cannot be derived solely from consistency requirements evident in low energy~\cite{Vafa:2005ui}. 
The absence of global symmetry is an example of such conditions. It has been long conjectured~\cite{Misner:1957mt,Polchinski:2003bq,Banks:2010zn} 
and recently proven in the context of the AdS/CFT correspondence~\cite{Harlow:2018tng,Harlow:2018jwu}
that any global symmetry in a low energy effective theory of consistent quantum gravity should be either broken or be a gauge symmetry in disguise\footnote{See also \cite{Banks:1988yz} for a worldsheet argument for the absence of continuous global symmetries. The recent work \cite{Harlow:2018tng,Harlow:2018jwu} provides a holographic proof of this conjecture and it covers additionally discrete symmetries as well as the completeness conjecture.}.
Since the launch of the Swampland program in 2005, a variety of Swampland conditions have been proposed with various degrees of rigors and 
motivations, and some of them have significant implications on cosmology and particle physics (see Refs.~\cite{Brennan:2017rbf,Palti:2019pca} for review articles).

\medskip
In this paper, we ask whether the Weak Gravity
Conjecture (WGC) can be derived from consistency conditions visible in low energy alone. 
Since the WGC is supposed to be stronger than the absence of global symmetry and since the proof of the latter requires~\cite{Harlow:2018tng,Harlow:2018jwu}
 knowledge on the microscopic mechanism of the AdS/CFT
correspondence such as the entanglement wedge reconstruction and its relation to quantum error correction\footnote{Readers who wish to learn about the relation can consult, for example,~\cite{Harlow:2018fse}.}, it is reasonable to expect that some information about UV physics is needed to prove the WGC. The purpose of this paper is to identify such UV information for a specific version of 
the WGC.

\medskip
The axionic WGC predicts the existence of instantons whose action-to-charge ratios are smaller than one in an appropriate unit~\cite{ArkaniHamed:2006dz}.
It connects the WGC to the distance conjecture~\cite{Ooguri:2006in} and 
imposes constraints on axion inflation scenarios\footnote{The axionic WGC constrains inflation scenarios with periodic axions, i.e., axions with a compact field space.  Axion monodromy inflation (using branes \cite{Silverstein:2008sg,McAllister:2008hb} and fluxes \cite{Marchesano:2014mla,Hebecker:2014eua,Blumenhagen:2014gta} to break the axion periodicity) provides an interesting exception, though other Swampland conditions can potentially constrain such models, see e.g. \cite{Palti:2019pca} for a review.} (see e.g. \cite{Brown:2015iha,Brown:2015lia,Montero:2015ofa, delaFuente:2014aca,Heidenreich:2015wga,Rudelius:2015xta,Junghans:2015hba,Bachlechner:2015qja, Hebecker:2016dsw,Hebecker:2017uix,Grimm:2019wtx,Heidenreich:2019bjd} and references therein)
and ultralight axion dark matter models ~\cite{Hebecker:2018ofv}. 
In this paper, we focus on the axion-gravity system and the axion-dilaton-gravity system.
We find that the WGC for the axion-gravity system follows from unitarity, analyticity, and locality of UV scattering amplitudes. On the other hand, these conditions are not sufficient for the axion-dilaton-gravity system; we find that the WGC for this system is satisfied if we in addition impose duality constraints.

\begin{figure}[t]
\begin{center}
\includegraphics[width=90mm, bb=0 0 438 213]{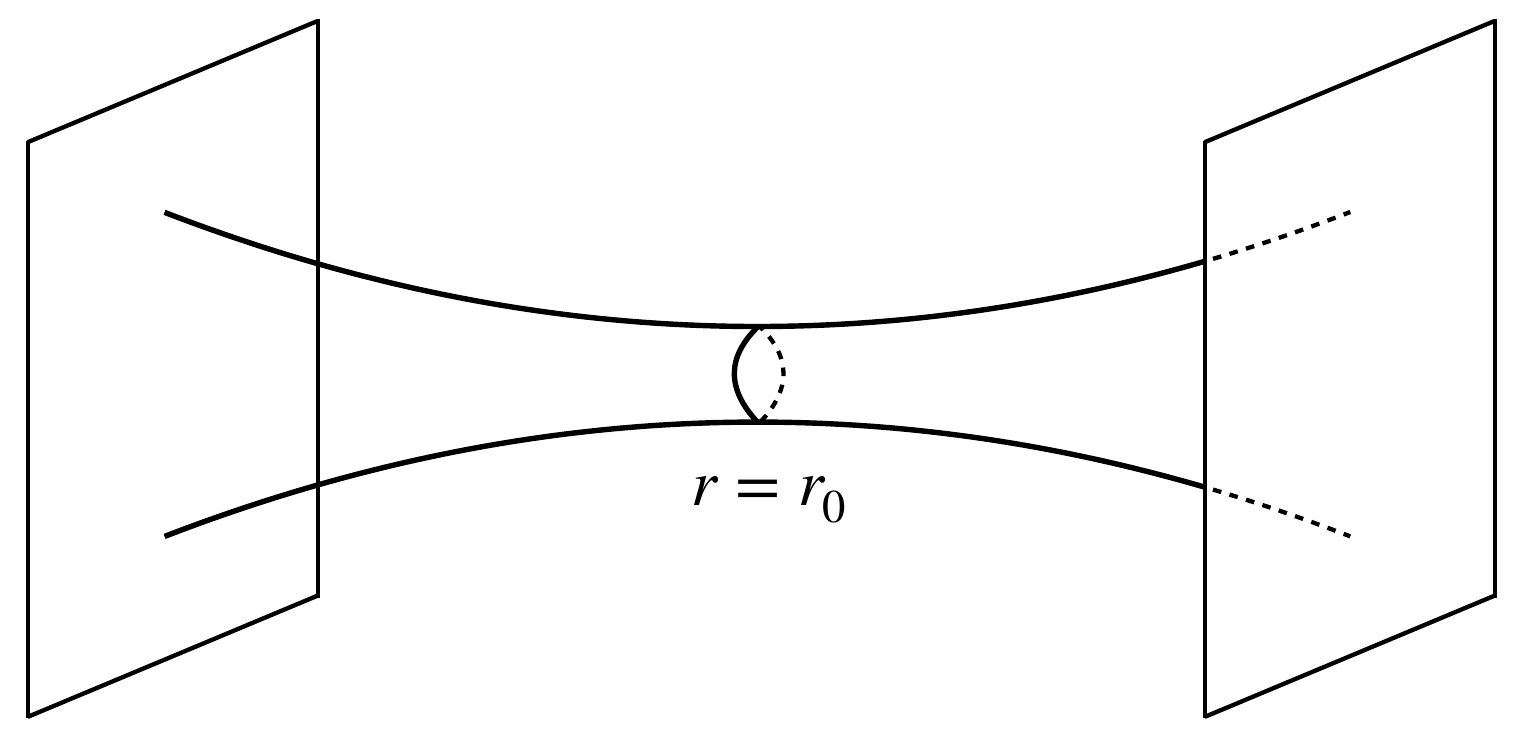}
\end{center}
\vspace{-5mm}
\caption{A wormhole connecting two asymptotically flat regions consists of two semiwormholes with opposite axion charges and the same action, which are glued at a three-sphere represented as $r=r_0$. Each semiwormhole can be regarded as an instanton.}
\label{wormhole}
\end{figure}

\medskip
In the 4D axion-gravity system, the upper bound is set by the action-to-charge ratio of the macroscopic semiwormhole (see Fig.~\ref{wormhole}) as\footnote{Since the notion of extremality for gravitational instantons is not clear (in contrast to the case for black branes), it is not fully understood yet how to formulate a precise version of the WGC for (-1)-form symmetries, see~\cite{Hebecker:2016dsw,Hebecker:2018ofv,VanRiet:2020pcn} and references therein. In this paper we follow~\cite{Hebecker:2016dsw,Hebecker:2018ofv} and use macroscopic wormholes, which are well controlled solutions in the EFT, as the reference to set the WGC bound.}
\begin{align}
\label{axion_WGC_no_dilaton}
\frac{S_n}{|n|}\leq \frac{\sqrt{6}\pi}{4}\cdot\frac{M_{\rm Pl}}{f}\,,
\end{align}
where $n$ and $S_n$ are the charge and action of the instanton required by the WGC, $M_{\rm Pl}$ is the reduced Planck mass, and $f$ is the axion decay constant. The WGC in this case guarantees that the tunneling process through a collection of small instantons dominates over the one through a single large instanton with the same charge. This is the axionic WGC counterpart of the statement ``every black hole has to decay" in the WGC for $0$-form symmetry.
Similarly, the WGC bound in the 4D axion-dilaton-gravity system is set by the action-to-charge ratio of the macroscopic semiwormhole as
\begin{align}
\label{axion_WGC}
\frac{S_n}{|n|}\leq \frac{2}{\beta}\sin\left[\tfrac{\sqrt{6}}{4}\beta\cdot\tfrac{\pi}{2}\right]\cdot\frac{M_{\rm Pl}}{f}\,,
\end{align}
where $\beta$ is the dilaton coupling defined shortly in Eq.~\eqref{Ead}. 
Throughout the paper we focus on the regime $|\beta|<\tfrac{4}{\sqrt{6}}$, where wormhole solutions exist and the bound~\eqref{axion_WGC} is applicable.
In the limit of $\beta \rightarrow 0$, when the dilaton decouples, \eqref{axion_WGC} reduces to \eqref{axion_WGC_no_dilaton}.

\medskip
The purpose of this paper is to clarify which properties of consistent UV completion are necessary for the WGC. To elaborate on this motivation, let us briefly review recent progress toward a proof of the WGC for $0$-form symmetry~\cite{Kats:2006xp,
Cheung:2014ega,
Nakayama:2015hga,Heidenreich:2015nta,Harlow:2015lma,
Benjamin:2016fhe,Heidenreich:2016aqi,Montero:2016tif,Cottrell:2016bty,
Crisford:2017gsb,Cheung:2018cwt,
Andriolo:2018lvp,Yu:2018eqq,Lee:2018urn,Hamada:2018dde,Lee:2018spm,Urbano:2018kax,Montero:2018fns,Bonnefoy:2018tcp,
Chen:2019qvr,Lee:2019tst,Bellazzini:2019xts,Aalsma:2019ryi,Heidenreich:2019zkl,Charles:2019qqt,Demirtas:2019lfi,Jones:2019nev,Loges:2019jzs,Goon:2019faz,Cano:2019oma,Cano:2019ycn,Montero:2019ekk,Cremonini:2019wdk,
Wei:2020bgk}. It has been known that higher derivative corrections to the Einstein-Maxwell theory modify the extremal condition of charged black holes and the macroscopic black holes play the role of the charged state required by the WGC, {\it if} the effective couplings satisfy a certain inequality~\cite{Kats:2006xp} (see Fig.~\ref{corrections}). The question has been which consistency conditions of the UV theory imply these inequality. In~\cite{Hamada:2018dde}, it was shown that the positivity bounds~\cite{Adams:2006sv} and the causality constraints~\cite{Camanho:2014apa} on higher derivative corrections imply the inequality and thus the WGC in a wide class of theories. However, more recently, \cite{Loges:2019jzs} studied the same problem in the photon-dilaton-gravity system to find that these consistency conditions are not enough to demonstrate the conjecture.

\begin{figure}[t]
\begin{center}
\includegraphics[width=60mm, bb=0 0 265 229]{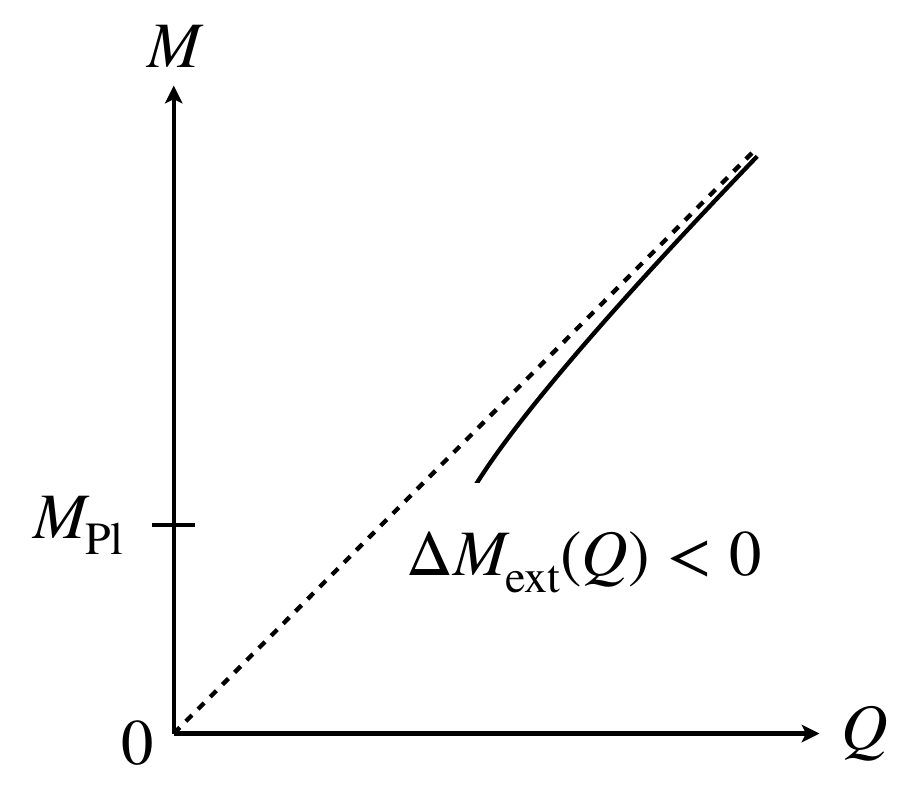}
\end{center}
\vspace{-5mm}
\label{corrections}
\caption{The dashed line represents the mass-charge relation of extremal black holes in the Einstein gravity, which defines the WGC bound.
If the higher derivative correction, $\Delta M_{\rm ext}(Q)$, to the mass of the extremal black hole with a fixed charge  $Q$ is negative (the solid curve), macroscopic black holes play the role of the charged state required by the WGC. This condition can be rephrased by a certain inequality of higher derivative couplings~\cite{Kats:2006xp,Natsuume:1994hd}. Similarly, the axionic WGC is satisfied if the higher derivative correction to the semiwormhole action is negative for a fixed charge.}
\label{corrections}
\end{figure}

\medskip
In this paper, we ask the same question for $(-1)$ form symmetry, namely instanton charges. We show that the positivity bound for  the axion-gravity system
imply the action-to-charge ratios \eqref{axion_WGC_no_dilaton} required by the WGC. On the other hand, for
the axion-dilaton-gravity system, the WGC does not follow from the positivity bounds alone.
Interestingly, the $SL(2,\mathbb{Z})$ symmetry\footnote{In \cite{Heidenreich:2016aqi,Montero:2016tif,Aalsma:2019ryi} , another $SL(2,\mathbb{Z})$ symmetry, i.e., the modular symmetry of the 2D CFT of the worldsheet (or the holographic dual) was used to upgrade the WGC to stronger forms. See Section \ref{Sec:discussion}.}
of the axion and dilaton, together with positivity bounds, does imply the axionic WGC. This presents an example where detailed UV information such as duality is needed to demonstrate the conjecture, on top of the positivity bounds which follow from unitarity, analyticity, and locality of UV scattering amplitudes.

\medskip
The rest of the paper is organized as follows: First, we evaluate the higher derivative correction to the semiwormhole solution and to its action (Sec.~\ref{Sec:correction}). Based on the results there, we discuss implications of positivity bounds and duality constraints to the axionic WGC (Sec.~\ref{Sec:WGC}). We conclude in Sec.~\ref{Sec:discussion} with discussion of our results. Some technical details in our derivations can be found in the Appendices.

\section{Higher derivative corrections to semiwormhole action}
\label{Sec:correction}

In this section we evaluate higher derivative corrections to the Euclidean action of the Giddings-Strominger (semi)wormhole.

\subsection{Giddings-Strominger wormhole}

We begin by a brief review of the Giddings-Strominger wormhole~\cite{Giddings:1987cg,Giddings:1989bq,GHS,Rey:1989xj} in the Einstein-axion-dilaton theory (see also \cite{Hebecker:2016dsw,Hebecker:2018ofv} for a review). Let us consider the Lorentzian action,
\begin{align}
\label{Ead}
S=\int \d^4 x \sqrt{-g}
\bigg[\frac{R}{2} - \frac{1}{2}(\de_\mu\phi)^2 - \frac{f^2}{2} e^{\beta\phi} (\de_\mu\theta)^2
 \bigg]\,,
\end{align}
where $\phi$ and $\theta$ are the dilaton and the axion, respectively, and we work in the unit $M_{\rm Pl}=1$. Without loss of generality, we assume that the dilaton coupling, $\beta$, is nonnegative: $\beta\geq0$. In this paper we focus on the regime, $0\leq\beta<\tfrac{4}{\sqrt{6}}$, so that the Giddings-Strominger wormhole exists. Also the dilaton may consistently be truncated (at the tree-level) when $\beta=0$ to reproduce the Einstein-axion theory.

\medskip
After Wick rotation, the Euclidean action corresponding to Eq.~\eqref{Ead} reads
\begin{align}
\label{Ead_Euclidean}
S_E=\int \d^4 x \sqrt{g}
\bigg[-\frac{R}{2} + \frac{1}{2}(\de_\mu\phi)^2 + \frac{f^2}{2} e^{\beta\phi} (\de_\mu\theta)^2
 \bigg]\,.
\end{align}
To work with wormhole solutions, it is convenient to dualize the axion into a two-form gauge field $B_{\mu\nu}$ as
\begin{align}
\label{Euclidean_Ead}
S_E=\int \d^4 x \sqrt{g}
\bigg[-\frac{R}{2} + \frac{1}{2}(\de_\mu\phi)^2 + \frac{1}{12f^2} e^{-\beta\phi} H_{\mu\nu\rho}^2
 \bigg]\,,
\end{align}
where $H=dB$ is the field-strength. The equations of motion are
\begin{align}
0&=\partial_\mu\left(\sqrt{g}e^{-\beta\phi}H^{\mu\nu\rho}\right)\,,
\\
0&=\frac{\beta\sqrt{g}}{12f^2}e^{-\beta\phi}H_{\mu\nu\rho}H^{\mu\nu\rho}+\partial_\mu\left(\sqrt{g}g^{\mu\nu}\partial_\nu\phi\right)\,,
\\
0&=R_{\mu\nu}-\partial_\mu\phi\partial_\nu\phi
+e^{-\beta\phi}\left(\frac{1}{6f^2}g_{\mu\nu}H_{\rho\sigma\lambda}H^{\rho\sigma\lambda}-\frac{1}{2f^2}H_{\mu\rho\sigma}H_\nu{}^{\rho\sigma}\right)\,.
\end{align}
The Giddings-Strominger solution is a spherically symmetric solution of the form,
\begin{align}
\label{GS_solution}
\d s^2=\frac{\d r^2}{1-\frac{r_0^4}{r^4}}+r^2\d\Omega_3^2\,,
\quad\,\,
H=\frac{n\varepsilon}{2\pi^2}\,,
\quad\,\,
e^{\beta\phi}=\frac{\cos^2\left[
\tfrac{\sqrt{6}}{4}\beta\cdot\arccos\frac{r_0^2}{r^2}
\right]}{\cos^2\left[
\tfrac{\sqrt{6}}{4}\beta\cdot\frac{\pi}{2}
\right]}\,,
\end{align}
where $\d\Omega_3^2$ and $\varepsilon$ are the line element and the volume form of a unit three-sphere, respectively, and the thickness, $r_0$, of the wormhole throat is given by
\begin{align}
r_0^4=\frac{n^2f^2}{24\pi^4}\cos^2\left[
\tfrac{\sqrt{6}}{4}\beta\cdot\tfrac{\pi}{2}
\right]
\,.
\end{align}
An explicit form of $\d\Omega_3^2$ and $\varepsilon$ is
\begin{align}
\d\Omega_3^2=\d\theta_1^2+\sin^2\theta_1\d \theta_2^2+\sin^2\theta_1\sin^2\theta_2\d\varphi^2\,,
\quad
\varepsilon=\sin^2\theta_1\sin\theta_2\d\theta_1\d\theta_2\d\varphi\,.
\end{align}

\medskip
The Giddings-Strominger solution~\eqref{GS_solution} has a coordinate singularity at $r=r_0$. The region, $r\geq r_0$, describes a half of the wormhole called the semiwormhole. We may identify $n$ with the axion charge of the semiwormhole:
\begin{align}
\int_{S^3} H=n\,.
\end{align}
The full wormhole connecting two asymptotically flat regions is obtained by gluing two such semiwormholes with opposite axion charges at a three-sphere defined by $r=r_0$  (see Fig.~\ref{wormhole}).
The on-shell action, $S_n$, of the semiwormhole~\eqref{GS_solution} with the charge $n$ is given by
\begin{align}
\label{GS_action}
S_n=
\int_{r_0}^\infty\d r\int_{S^3}\d^3x\sqrt{g}
\left[\frac{1}{6f^2}e^{-\beta\phi}H_{\mu\nu\rho}^2\right]
=\frac{2|n|}{\beta f}\sin\left[\tfrac{\sqrt{6}}{4}\beta\cdot\tfrac{\pi}{2}\right]\,,
\end{align}
where the integral region is $r_0\leq r<\infty$ because we are considering the semiwormhole action. The action of the full wormhole is obtained by multiplying by a factor of $2$.
Note that we have omitted boundary terms such as the Gibbons-Hawking-York term. However, it is easy to show that such boundary terms have vanishing contributions to the wormhole action in our analysis. See also Appendix~\ref{app:dual}. Now the action-to-charge ratio of the macroscopic semiwormhole is
\begin{align}
\frac{S_n}{|n|}=\frac{2}{\beta}\sin\left[\tfrac{\sqrt{6}}{4}\beta\cdot\tfrac{\pi}{2}\right]\cdot\frac{M_{\rm Pl}}{f}\,,
\end{align}
which sets the WGC bound~\eqref{axion_WGC} on the action-to-charge ratio of the required instanton.

\subsection{Higher derivative corrections}

Next let us consider higher derivative corrections to the Einstein-axion-dilaton theory:    
\begin{align}
\label{axion_higher_derivative_Lorentzian}
S=S^{(0)}+\Delta S\,,
\end{align}
where $S^{(0)}$ is the Einstein-axion-dilaton action~\eqref{Ead} and $\Delta S$ is for higher derivative terms. At the leading order, the general parity-even operators which respect the axion shift symmetry are given by\footnote{Note that the continuous axion shift symmetry may be broken to a discrete one by non-perturbative effects. Since such non-perturbative effects are exponentially suppressed, we assume that the dominant contribution in our analysis is captured by the continuous shift symmetric terms presented here.
}
\begin{align}
\Delta S&=\int d^4x \sqrt{-g}
\Big[\,
a_1(\phi)\,(\partial_\mu\phi\partial^\mu\phi)^2
+a_2(\phi)\,f^4(\partial_\mu\theta\partial^\mu\theta)^2
\nonumber
\\
&\qquad\qquad\qquad\,\,
+a_3(\phi)\,f^2(\partial_\mu\phi\partial^\mu\phi)(\partial_\mu\theta\partial^\mu\theta)
+a_4(\phi)\,f^2(\partial_\mu\phi\partial^\mu\theta)^2
\nonumber
\\
&\qquad\qquad\qquad\,\,
+a_5(\phi)\,W_{\mu\nu\rho\sigma}^2
+a_6\,\theta W_{\mu\nu\rho\sigma}\widetilde{W}^{\mu\nu\rho\sigma}
\,\Big]\,,
\label{higher_derivative_axion}
\end{align}
where
$W_{\mu\nu\rho\sigma}$ is the Weyl tensor and $\widetilde{W}
_{\mu\nu\rho\sigma}=\frac{1}{2}\epsilon_{\mu\nu}{}^{\alpha\beta}W_{\alpha\beta\rho\sigma}$. For generality, we assumed that the EFT coefficients $a_{i}$ ($i=1,\ldots , 5$) are general functions of $\phi$ ($\phi$-dependence of $a_6$ is prohibited by the axion shift symmetry).
Also note that
we performed field redefinition to eliminate unphysical degeneracy of the effective operators.
In the two-form language, the corresponding Euclidean action reads (see Appendix~\ref{app:dual} for details of dualization)
\begin{align}
\label{Euclidean_two_form_action}
S_E=S_E^{(0)}+\Delta S_E\,,
\end{align}
where $S_E^{(0)}$ is the Einstein-axion-dilaton theory one~\eqref{Euclidean_Ead} and
\begin{align}
\Delta S_E&=\int d^4\sqrt{g}
\left[
-a_1(\phi)\,(\partial_\mu\phi\partial^\mu\phi)^2
-\frac{a_2(\phi)}{36f^4}e^{-4\beta\phi}(H_{\mu\nu\rho}H^{\mu\nu\rho})^2
\right.
\nonumber
\\
&\qquad\qquad\quad\,\,
+\frac{a_3(\phi)}{6f^2}e^{-2\beta\phi}(\partial_\mu\phi\partial^\mu\phi)(H_{\mu\nu\rho}H^{\mu\nu\rho})
+\frac{a_4(\phi)}{36f^2}e^{-2\beta\phi}(\epsilon^{\mu\nu\rho\sigma}\partial_\mu\phi H_{\nu\rho\sigma})^2
\nonumber
\\
&\qquad\qquad\quad\,\,
\left.
-a_5(\phi)\,W_{\mu\nu\rho\sigma}^2
-\frac{a_6}{6}H_{\mu\nu\rho}J^{\mu\nu\rho}
\right]\,.
\label{higher_derivative_two-form}
\end{align}
Here we introduced a three-form
$J$ satisfying $W_{\mu\nu\rho\sigma}\widetilde{W}^{\mu\nu\rho\sigma}=-\star dJ$ which is dual to the Chern-Simons current. Its explicit form is given by
\begin{align}
\label{def_J}
J_{\mu\nu\rho}&=
-12\left(\Gamma^\alpha_{\beta\mu}\partial_\nu\Gamma^\beta_{\alpha\rho}+\frac{2}{3}\Gamma^\alpha_{\beta\mu}\Gamma^\beta_{\gamma\nu}\Gamma^\gamma_{\alpha\rho}\right)+\text{5 similar terms}
\,.
\end{align}
Note that the three-form $J$ is not covariant under coordinate transformations, but its exterior derivative $\d J$ is covariant. Also $J$ vanishes in conformally flat coordinates.

\medskip
We now evaluate the higher derivative corrections to the semiwormhole action. Since the equations of motion are modified by the higher derivative terms, the wormhole solutions are also modified. We write the modified solution schematically as
\begin{align}
\Phi=\Phi^{(0)}+\Delta \Phi\,,
\end{align}
where $\Phi$ stands for all the fields $g_{\mu\nu},B_{\mu\nu},\phi$. $\Phi^{(0)}$ is the wormhole solution~\eqref{GS_solution} in the Einstein-axion-dilaton theory and $\Delta\Phi$ is the higher derivative correction. Note that the two-form field solution, $\displaystyle H=\frac{n\varepsilon}{2\pi^2}$, in Eq.~\eqref{GS_solution} is not corrected by higher derivative terms because of charge quantization (recall that the wormhole solution has a two-form magnetic charge).
The leading order correction to the semiwormhole action is then given by
\begin{align}
\Delta S_n&=S_E[\Phi]-S_E^{(0)}[\Phi^{(0)}]
\nonumber
\\
&=\left(S_E^{(0)}[\Phi^{(0)}+\Delta\Phi]-S_E^{(0)}[\Phi^{(0)}]\right)+\Delta S_E[\Phi^{(0)}+\Delta \Phi]
\nonumber
\\
&=\Delta S_E[\Phi^{(0)}]+\mathcal{O}(\Delta^2)\,,
\end{align}
where $\mathcal{O}(\Delta^2)$ stands for higher order corrections. Note that at the leading order, the first term in the second line is proportional to the equations of motion up to a boundary term. One may explicitly show that the boundary term vanishes in our wormhole analysis, so that the first term has no leading order contribution. Our task is now to evaluate
\begin{align}
\label{correction_action_implicit}
\Delta S_n&=
\int_{r_0}^\infty dr\int_{S^3}\d^3x\sqrt{g}
\left[
-a_1(\phi)\,(\partial_\mu\phi\partial^\mu\phi)^2
-\frac{a_2(\phi)}{36f^4}e^{-4\beta\phi}(H_{\mu\nu\rho}H^{\mu\nu\rho})^2
\right.
\nonumber
\\
&\qquad\qquad\qquad\qquad\quad
+\left.
\frac{a_3(\phi)+a_4(\phi)}{6f^2}e^{-2\beta\phi}(\partial_\mu\phi\partial^\mu\phi)(H_{\mu\nu\rho}H^{\mu\nu\rho})
\right]\,,
\end{align}
where we used the fact that the wormhole solution~\eqref{GS_solution} is conformally flat (see discussion around Eq.~\eqref{def_J})
and it satisfies the relation, $(\epsilon^{\mu\nu\rho\sigma}\partial_\mu\phi H_{\nu\rho\sigma})^2=6(\partial_\mu\phi\partial^\mu\phi)(H_{\mu\nu\rho}H^{\mu\nu\rho})$.
More explicitly, we find
\begin{align}
\nonumber
\Delta S_n&=36\pi^2\int_0^{\frac{\pi}{2}}\d t\cos^3t
\bigg[
-a_1\big(\phi(t)\big)\tan^4\Big[\tfrac{\sqrt{6}}{4}\beta\cdot t\Big]
-a_2\big(\phi(t)\big)e^{-2\beta\phi(t)}\sec^4\Big[\tfrac{\sqrt{6}}{4}\beta\cdot t\Big]
\\
\label{correction_action}
&\qquad\qquad\qquad
+\Big(a_3\big(\phi(t)\big)
+a_4\big(\phi(t)\big)\Big)e^{-\beta\phi(t)}\tan^2\Big[\tfrac{\sqrt{6}}{4}\beta\cdot t\Big]\sec^2\Big[\tfrac{\sqrt{6}}{4}\beta\cdot t\Big]\,
\bigg]\,,
\end{align}
where $\phi(t)$ is defined such that
\begin{align}
e^{\phi(t)}=\frac{\cos^2\Big[\tfrac{\sqrt{6}}{4}\beta\cdot t\Big]}{\cos^2\Big[\tfrac{\sqrt{6}}{4}\beta\cdot\frac{\pi}{2}\Big]}\,.
\end{align}
It is difficult to evaluate the integral analytically for general $a_i$ and $\beta$, but one can explicitly check that it is finite in the regime, $0\leq\beta<\tfrac{4}{\sqrt{6}}$, of our interests, as long as the EFT coefficients $a_i(\phi)$ are regular throughout the wormhole geometry. We find that the prefactor of $a_{1,2}$ is always negative, whereas that of $a_{3,4}$ is positive. Also note that the correction~\eqref{correction_action} to the action is $n$-independent at this order and thus the correction to the action-to-charge ratio is suppressed by $1/n$, which guarantees validity of the derivative expansion for macroscopic wormholes with a sufficiently large charge.

\section{Axionic WGC v.s. positivity bounds}
\label{Sec:WGC}

In the previous section we evaluated the higher derivative correction, $\Delta S_n$, to the action of the semiwormhole with a fixed charge $n$.
The axionic WGC in the original form is then satisfied by macroscopic semiwormholes if the correction~\eqref{correction_action} is negative and therefore the action-to-charge ratio decreases for a fixed charge. In this section we demonstrate that the axionic WGC follows from the duality constraints together with the positivity bounds on higher derivative couplings. On the other hand, it does not follow from the positivity bounds alone when the dilaton coupling $\beta$ is nonzero.

\subsection{Positivity bounds are not enough to imply axionic WGC}

\paragraph{Axion-gravity system}

Let us first discuss implications of positivity bounds in the axion-gravity system without the dilaton ($a_2$ has no $\phi$-dependence in particular):
\begin{align}
S=\int \d^4x\sqrt{-g}\left[
\frac{R}{2}-\frac{f^2}{2}(\partial_\mu\theta)^2+a_2f^4(\partial_\mu\theta\partial^\mu\theta)^2
\right]\,.
\end{align}
It is known that unitarity, analyticity, and locality of UV scattering amplitudes imply the positivity bound: $a_2\geq0$~\cite{Adams:2006sv}.
To be precise, the positivity bounds on the four-derivative operator can be proven only when gravity is negligible~\cite{Adams:2006sv}. More quantitatively, if we assume that gravity is UV completed by an infinite higher-spin Regge tower, the bound is applicable when $|a_{i}|\gg (M_{\rm Pl}^2M_{\rm Regge}^2)^{-1}$. Here $M_{\rm Regge}$ is the mass scale of the gravitational Regge states, i.e., the string scale $M_s$ in string theory. See Appendix I of Ref.~\cite{Hamada:2018dde} for details. In this paper we assume that there exists such a hierarchy and discuss implications of positivity bounds.
On the other hand, the higher derivative correction to the semiwormhole action is given by Eq.~\eqref{correction_action} with $\beta=0$, for which we may evaluate the integral analytically to obtain
\begin{align}
\Delta S_n=-24\pi^2a_2\,.
\end{align}
We find that the positivity bound $a_2\geq0$ directly implies the negativity of the correction to the semiwormhole action and thus the axionic WGC.

\paragraph{Axion-dilaton-gravity system}

We then incorporate the dilaton. As we discussed, there are four higher derivative operators $a_{1,2,3,4}$ relevant to the leading order correction to the semiwormhole action. Positivity bounds for $\phi\phi\to\phi\phi$, $\theta\theta\to\theta\theta$, and $\phi\theta\to\phi\theta$ scattering imply that 
three of them are positive:
\begin{align}
a_1(\phi)\geq 0\,, \quad a_2(\phi)\geq 0\,,\quad a_4(\phi)\geq 0\,,
\end{align}
which should hold for all $\phi$ since we may consider scattering around arbitrary constant dilaton backgrounds (see also Appendix~\ref{Positivity-bounds}).
More generally, by considering scattering of the superposition of the axion and the dilaton, we may derive a family of positivity bounds~\cite{Andriolo:2018lvp}
\begin{align}
\label{family_positivity}
u_1^2v_1^2\,a_1(\phi)+u_2^2v_2^2\,a_2(\phi)+u_1u_2v_1v_2\,a_3(\phi)+\frac{1}{4}(u_1v_2+u_2v_1)^2\,a_4(\phi)\geq0\,,
\end{align}
where $u_i$ and $v_i$ are arbitrary real numbers characterizing the mixing of the dilaton and the axion in the external states. 
The allowed region of the higher derivative operators consistent with the positivity bounds is thus:
\begin{align}
\label{positivity_final}
& a_1\geq 0\,,
\quad
a_2\geq 0\,,
\quad
a_4\geq 0\,, 
\quad
-a_4 - 2 \sqrt{a_1 a_2} \leq a_3 \leq  2 \sqrt{a_1 a_2}\,,
\end{align}
where $\phi$-dependence is implicit for notational simplicity. See also Appendix \ref{Positivity-bounds} for a derivation of the bounds.

\medskip
Now let us get back to the correction~\eqref{correction_action}. First, positivity of $a_1$ and $a_2$ implies that these operators decrease the semiwormhole action as expected by the WGC. On the other hand, the $a_4$ operator always increases the action as a consequence of positivity! It is counterintuitive from the WGC perspective. Indeed, there exists a certain parameter space which is consistent with the bounds~\eqref{positivity_final}, but the correction to the semiwormhole action is positive (an extreme example is  $a_4\gg a_1,a_2,|a_3|$ with $a_1,a_2>0$ and $a_3<0$). Therefore, in order for macroscopic semiwormholes to play the role of the instanton required by the axionic WGC, $a_{1,2}$ have to be large enough to compensate the positive contribution from the $a_4$ operator (and the $a_3$ operator if $a_3>0$). This illustrates a situation where the positivity bounds are not powerful enough to imply the axionic WGC.

\subsection{Implications of duality constraints}

While the positivity bounds are universal UV constraints on the low-energy effective theory, they are a  {\it subset} of the necessary conditions for the low-energy theory to have a consistent UV completion. Indeed, we know that stronger constraints can be obtained once we specify more details of the UV completion. Therefore, it is important to clarify which ingredients of UV information, together with the positivity bound, lead to the axionic WGC. Below, we show that the $SL(2,\mathbb{Z})$ symmetry of the axion and dilaton may play the role\footnote{
Note that the $SL(2,\mathbb{Z})$ invariance is not a necessary condition for consistent UV completion of quantum gravity. For example, heterotic string is not self-dual under the $S$-duality transformation, so that the $\alpha'$ corrections for the string dilaton and the universal axion (dual to the NS-NS two-form) do not satisfy the $SL(2,\mathbb{Z})$ invariance, even though the dilaton and axion with other origins may respect the invariance. Our purpose here is to provide an example for the UV condition which guarantees the WGC, rather than to complete the full list of the UV scenarios which lead to the conjecture.}.

\medskip
In our convention, the $SL(2,\mathbb{Z})$ transformation is defined by
\begin{align}
\tau\to\frac{a\tau+b}{c\tau+d}
\quad
(a,b,c,d\in \mathbb{Z}, \,\,ad-bc=1)\,
\quad
{\rm with}
\quad
\tau=\frac{\beta }{2}f\theta+ie^{-\frac{\beta}{2}\phi}\,.
\end{align}
For example, the kinetic terms of the dilaton and axion can be recast as
\begin{align}
-\frac{1}{2}(\partial_\mu\phi)^2-\frac{f^2}{2}e^{\beta\phi}(\partial_\mu\theta)^2=-\frac{1}{2}\frac{\partial_\mu\tau\partial^\mu\bar{\tau}}{\left(\frac{\beta}{2}\right)^2({\rm Im}\tau)^2}\,,
\end{align}
which manifests the $SL(2,\mathbb{Z})$ invariance. Similarly, $SL(2,\mathbb{Z})$ invariant four-derivative operators are of the form,
\begin{align}
\lambda_1\frac{(\partial_\mu\tau\partial^\mu\bar{\tau})^2}{\left(\frac{\beta}{2}\right)^4({\rm Im}\tau)^4}
+\lambda_2\frac{(\partial_\mu\tau\partial^\mu\tau)(\partial_\mu\bar{\tau}\partial^\mu\bar{\tau})}{\left(\frac{\beta}{2}\right)^4({\rm Im}\tau)^4}\,,
\end{align}
where $\lambda_{1,2}$ are ($\phi$-independent) constants. This corresponds to the parameter set,
\begin{align}
a_1=\lambda_1+\lambda_2\,,
\quad
a_2=(\lambda_1+\lambda_2)e^{2\beta\phi}\,,
\quad
a_3=2(\lambda_1-\lambda_2)e^{\beta\phi}
\,,
\quad
a_4=4\lambda_2e^{\beta\phi}\,.
\end{align}
Also the positivity bounds~\eqref{positivity_final} are reduced to $\lambda_1+\lambda_2\geq0$ and $\lambda_2\geq0$.

\medskip
Then, in the $SL(2,\mathbb{Z})$ invariant theory, the integral in Eq.~\eqref{correction_action} drastically simplifies to obtain the higher derivative correction to the semiwormhole action,
\begin{align}
\Delta S_n
&=-36\pi^2 (\lambda_1+\lambda_2) \int_0^{\frac{\pi}{2}}\d t\cos^3t\left(\tan^2t-\sec^2t\,\right)^2
\nonumber
\\
&=-36\pi^2 (\lambda_1+\lambda_2) \int_0^{\frac{\pi}{2}}\d t\cos^3t
=-24\pi^2 (\lambda_1+\lambda_2)
\,.
\end{align}
Therefore, the negativity of the correction~\eqref{correction_action} and thus the axionic WGC follows from the positivity bound $\lambda_1+\lambda_2\geq0$. To summarize, the $SL(2,\mathbb{Z})$ invariance together with the positivity bounds implies the axionic WGC, even though the conjecture does not follow if we assume the positivity bounds only. This presents an example where detailed UV information such as duality is needed to demonstrate the conjecture.

\section{Discussion}
\label{Sec:discussion}

In this paper we showed that the WGC for $(-1)$ form symmetry in the axion-gravity system 
follows from the positivity bounds on higher derivative corrections and that in the axion-dilaton-gravity system
it also follows if we in addition assume certain duality constraints. 
Our work provides strong evidences for the axionic WGC, which has significant implications in 
axion inflation scenarios and ultralight axion dark matter models.
At the same time, it also
illustrates examples of UV ingredients that are necessary for Swampland conditions to hold.
In particular, we find necessary UV ingredients depend on the field contents in the IR. 
It is our hope that research in this direction will help progress toward proofs of the Swampland criteria. 
In view of this, it is important to clarify how crucial duality is in demonstrating the WGC or if there are other UV
ingredients that also imply it.
It would also be interesting to generalize our results to the WGC for $p (\geq 0)$-form symmetry (See
 Ref.~\cite{Loges:2019jzs}).

\medskip
It is also worth mentioning a connection to stronger versions of WGC such as the sublattice/tower WGC~\cite{Heidenreich:2015nta,Heidenreich:2016aqi,Montero:2016tif,Andriolo:2018lvp}. For example, Refs.~\cite{Heidenreich:2016aqi,Montero:2016tif,Aalsma:2019ryi} argued that in the presence of a heavy state satisfying the WGC bound, the notion of spectral flow (which follows from the modular invariance in the 2D CFT of the worldsheet or the holographic dual) implies a tower of states below the Planck scale satisfying the WGC bound. In other words, we may upgrade the original WGC to stronger versions by adding one more UV information that the UV theory has a worldsheet structure of the perturbative string theory or a holographically dual 2D CFT. It appears that a stronger version of the conjecture follows from a weaker version by assuming more details of the UV completion. Further research in this direction may
provide a global view on the web of Swampland conditions and how different conditions reflect different UV ingredients.

\section*{Acknowledgments}
We would like to thank Yu-tin Huang and Pablo Soler for valuable discussions and comments on the earlier version of the draft. 
S.~A.\ is supported by ``Fondazione Angelo Della Riccia'' Fellowship.
T.~N.\ is supported in part by JSPS KAKENHI Grant
Numbers JP17H02894, JP18K13539 and JP20H01902, and MEXT KAKENHI Grant Number
JP18H04352. 
The work of H.~O. is supported in part by
U.S.\ Department of Energy grant DE-SC0011632, by
the World Premier International Research Center Initiative,
MEXT, Japan, by JSPS Grant-in-Aid for Scientific Research C-26400240,
and by JSPS Grant-in-Aid for Scientific Research on Innovative Areas
15H05895.
 G.~S.\ is supported in part by the DOE grant DE-SC0017647
and the Kellett Award of the University of Wisconsin.
T.~N.\ and G.~S.\ gratefully acknowledge the hospitality of the Kavli Institute for Theoretical Physics (supported by 
NSF PHY-1748958) while part of this work was completed.
H.~O. thanks the Aspen Center for Theoretical Physics, which is supported by
the National Science Foundation grant PHY-1607611,  where part of this work was done.

\appendix

\section{Dualization}
\label{app:dual}

In this appendix we provide details of the dualization procedure translating the axion description into the two-form description. To elaborate on the boundary conditions in our problem and the corresponding boundary terms, we begin by the Einstein-axion-dilaton theory without higher derivative terms.

\paragraph{Einstein-axion-dilaton theory}

To describe the dualization procedure in our wormhole analysis, let us consider the following parent action in the Euclidean signature:
\begin{align}
\label{parent_leading}
S_E &= \int\d^4 x\sqrt{g}\left[
-\frac{R}{2}+\frac{1}{2}(\de_\mu\phi)^2
\right]  - S_{GHY}
+ \int
\left[
\frac{1}{2f^2}
e^{-\beta\phi}
H\wedge \star H
+
i\theta \d H
\right]
\,,
\end{align}
where $S_{GHY}$ is the Gibbons-Hawking-York term and the three-form field, $H$, is unconstrained at this moment. Integrating out $\theta$, we have $H=\d B$ with $B$ being a two-form field to reproduce Eq.~\eqref{Euclidean_Ead}. Since semiwormholes have a magnetic charge for the two-form gauge field, we impose boundary conditions at asymptotic infinity on the three-form $H$ such that they are consistent with the charge quantization. More explicitly, we impose $\displaystyle H=\frac{n\varepsilon}{2\pi^2}$ ($n \in \integers$), where $ \epsilon$ is the volume form of a unit three-sphere. For these boundary conditions, the action~\eqref{parent_leading} gives a consistent variation problem without requiring any additional boundary term.

\medskip
On the other hand, the action for the axion $\theta$ is obtained by integrating out $H$. To do so, we rewrite the last term in Eq.~\eqref{parent_leading} as
\begin{align}
\int
\left[
\frac{1}{2f^2}
e^{-\beta\phi}
\big(H-i f^2e^{\beta\phi}\star \d\theta\big)\wedge \star \big(H-i f^2e^{\beta\phi}\star \d\theta\big)
+
\frac{f^2}{2}e^{\beta\phi}\d\theta \wedge \star \d\theta
+i\d(\theta H)
\right]
\,.
\end{align}
Integrating out $H$ reproduces the action~\eqref{Ead_Euclidean} up to a total derivative term, i.e., the last term shown in the above. The corresponding boundary term, $\int_{\partial\mathcal{M}} i\theta H$, is nothing but the one required to make the variation problem consistent and describe transitions between eigenstates of axion charge (see also Refs.~\cite{Collinucci:thesis,ArkaniHamed:2007js,Hebecker:2016dsw} for details): it has to be appropriately taken into account when evaluating the on-shell action in the axion language. Note that the boundary term does not participate in integrating out $H$ because its boundary value is fixed by the boundary conditions. Also note that the axion $\theta$ has a pure imaginary configuration for wormhole solutions, which means that wormhole solutions are complex saddle points in the axion path integral. The two descriptions are equivalent, but we employed the two-form language in the main text for computational simplicity.

\paragraph{Higher derivative corrections}

Next we incorporate higher derivative corrections. Recall that higher derivative corrections to the wormhole solution and the boundary action decay quickly far away from the wormhole neck\footnote{Note that boundary terms are located only at asymptotic infinity. In particular, there are no boundary terms at the boundary $r=r_0$ of the two semiwormholes, essentially because it is a coordinate singularity which can be eliminated by coordinate transformations, just as the black hole horizon.}. Then, the above argument on boundary conditions and boundary terms is unchanged even in the presence of higher derivative operators. We therefore focus on the duality transformation of the bulk action.

\medskip
Similarly to the previous case, let us consider the following parent action:
\begin{align}
\label{parent_corrections}
S_E &= \int\d^4 x\sqrt{g}\left[
-\frac{R}{2}+\frac{1}{2}(\de_\mu\phi)^2
\right]
+ \int
\left[
\frac{1}{2f^2}
e^{-\beta\phi}
H\wedge \star H
+
i\theta \d H
+\alpha(H,\phi,g_{\mu\nu})
\right]
\,,
\end{align}
where $\alpha(H,\phi,g_{\mu\nu})$ denotes collectively all the four-derivative operators in Eqs.~\eqref{Euclidean_two_form_action}-\eqref{higher_derivative_two-form} with a replacement of $H=dB$ by an unconstrained three-form $H$. Also we neglect total derivatives as mentioned. First, we reproduce the action~\eqref{Euclidean_two_form_action}-\eqref{higher_derivative_two-form} by integrating out~$\theta$. On the other hand, the equation of motion for $H$ reads $H=if^2e^{2\beta\phi}\star \d\theta+\mathcal{O}(\alpha)$, where $\mathcal{O}(\alpha)$ denotes higher derivative corrections. Substituting this into Eq.~\eqref{parent_corrections}, we obtain
\begin{align}
S_E &= \int\d^4 x\sqrt{g}\left[
-\frac{R}{2}+\frac{1}{2}(\de_\mu\phi)^2
\right]
+ \int
\left[
\frac{f^2}{2}
e^{\beta\phi}
\d\theta\wedge \star \d\theta
+\alpha(if^2e^{2\beta\phi}\star \d\theta,\phi,g_{\mu\nu})
\right]
\end{align}
up to six-derivative operators and higher orders. This gives the axion action~\eqref{axion_higher_derivative_Lorentzian}-\eqref{higher_derivative_axion} after Wick rotation.

\section{Positivity bounds}
\label{Positivity-bounds}

In this appendix we provide a derivation of the bounds~\eqref{positivity_final} applying the argument in~\cite{Andriolo:2018lvp}. Let us consider four-point forward scattering of the dilaton and the axion around the background, $g_{\mu\nu}=\eta_{\mu\nu}$, $\theta=0$ and $\phi=\phi_*$ (constant), which is realized, e.g., at the asymptotic infinity far away from the wormhole neck. Note that $\phi_*$ can be chosen as an arbitrary number, as long as the axioin-dilaton coupling is in the perturbative regime. Up to the four-derivative order in the low-energy expansion, nonzero forward amplitudes in our setup~\eqref{axion_higher_derivative_Lorentzian} are given by
\begin{align}
\mathcal{M}_{\phi\phi\to\phi\phi}&=4a_1s^2\,,
\\
\mathcal{M}_{\theta\theta\to\theta\theta}&=4a_2s^2\,,
\\
\mathcal{M}_{\phi\theta\to\phi\theta}&=\mathcal{M}_{\theta\phi\to\theta\phi}=a_4s^2\,,
\\
\mathcal{M}_{\phi\phi\to\theta\theta}&=\mathcal{M}_{\theta\theta\to\phi\phi}=-\frac{\beta^2}{8}s+\left(a_3+\frac{a_4}{2}\right)s^2\,,
\\
\mathcal{M}_{\phi\theta\to\theta\phi}&=\mathcal{M}_{\theta\phi\to\phi\theta}=\frac{\beta^2}{8}s+\left(a_3+\frac{a_4}{2}\right)s^2\,,
\end{align}
where, e.g., $\mathcal{M}_{\phi\phi\to\phi\phi}$ denotes $\phi\phi\to\phi\phi$ scattering in the forward limit and $a_i$ are evaluated at $\phi=\phi_*$. Also in our analysis we neglect graviton exchange diagrams, which can be justified at least when gravity is UV completed by an infinite higher-spin Regge tower and the gravitational Regge states are subdominant contributions to the $s^2$ coefficient~\cite{Hamada:2018dde} 
.

\medskip
Next we apply the positivity bound on the $s^2$ coefficient. Recall that the positivity argument in~\cite{Adams:2006sv} is applicable only to $s$-$u$ symmetric amplitudes. To parameterize such $s$-$u$ symmetric amplitudes, let us write
\begin{align}
\mathcal{M}(u_i,v_i)=\sum_{i,j,k,l}u_iv_ju_kv_l\mathcal{M}_{ij\to k\ell}\quad
(i,j,k,\ell=\phi,\theta)\,,
\end{align}
where $u_i$ and $v_i$ are arbitrary real numbers characterizing the mixing of the dilaton and the axion in the external states. Then, the positivity of the $s^2$ coefficient of the amplitude $\mathcal{M}(u_i,v_i)$ implies a family of positivity bounds~\cite{Andriolo:2018lvp}:
\begin{align}
\label{positivity1}
u_1^2v_1^2a_1+u_2^2v_2^2a_2+u_1v_1u_2v_2a_3+\frac{1}{4}(u_1v_2+u_2v_1)^2a_4\geq0\,.
\end{align}
For example, specific parameter sets, $(u_1,v_1,u_2,v_2)=(1,1,0,0),(0,0,1,1),(1,0,0,1)$, reproduce the positivity bounds for $\phi\phi\to\phi\phi$, $\theta\theta\to\theta\theta$, and $\phi\theta\to\phi\theta$ scattering:
\begin{align}
\label{124_positivity1}
a_1\geq0\,,\quad
a_2\geq0\,,\quad
a_4\geq0\,.
\end{align}

\begin{figure}[t]
	\includegraphics[scale=.5]{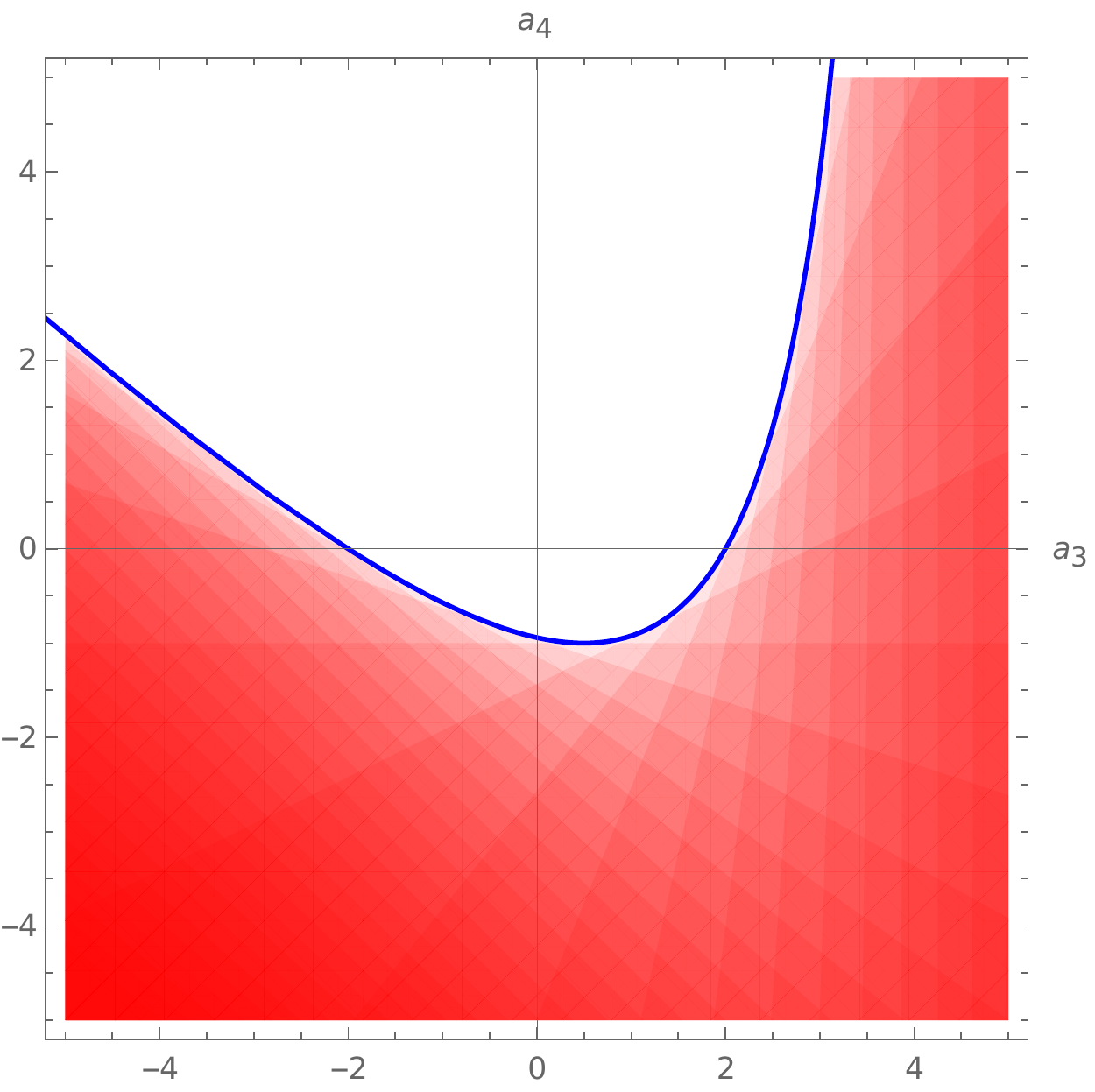}\qquad
	\includegraphics[scale=.5]{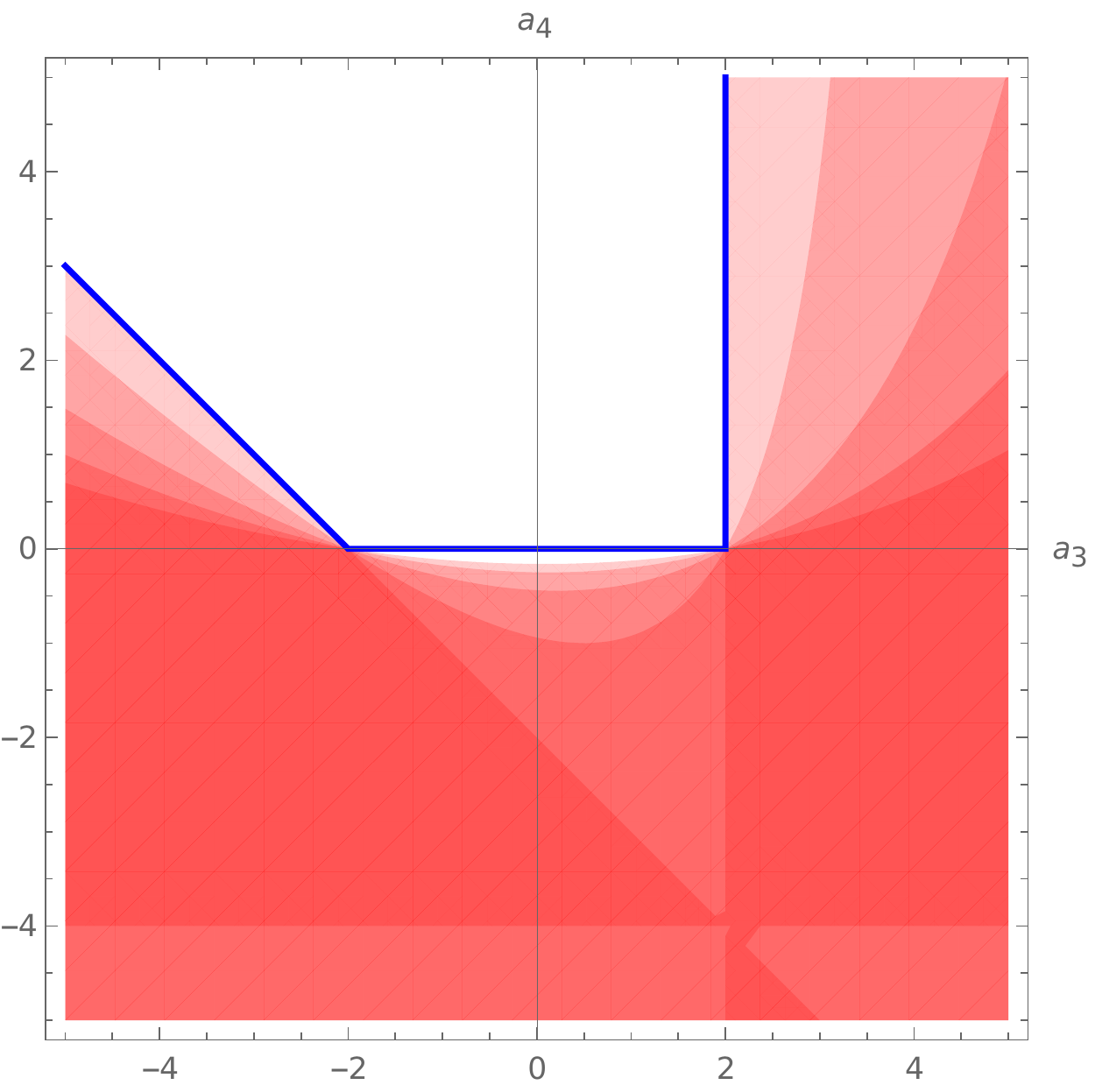}
	\centering
	\caption{
Positivity bounds in the $a_1=a_2=1$ plane as envelope of~\eqref{positivity1}: Without loss of generality, let us take $u_2=v_2=1$. Then, the bounds~\eqref{positivity1} for a fixed $v_1$ give a one-parameter family of linear relations of $a_i$ parameterized by $u_1$. Varying $u_1$ and taking the envelop, we obtain a nonlinear relation for each $v_1$. For example, the left figure is for $v_1=1$, where the straight lines represent the linear relation for each $u_1$ and the red region is excluded. The blue curve is the envelop defining the nonlinear relation. Varying the resulting envelop in $v_1$, we reproduce the bound~\eqref{positivity_final}, represented by the white region surrounded by the blue lines in the right figure (each red curve represents the envelop for a fixed $v_1\in [0,10]$). It is also straightforward to derive the bound~\eqref{positivity_final} analytically by this prescription.}
\label{Fig:envelop}
\end{figure}

\medskip
Finally, we show that the family of positivity bounds~\eqref{positivity1} is equivalent to Eq.~\eqref{positivity_final}. Since the inequality~\eqref{positivity1} is invariant under sign flip of any pair of the four parameters, $u_{1,2}$ and $v_{1,2}$, e.g., $u_{1,2}\to-u_{1,2}$, we assume $u_1,u_2,v_1\geq0$ without loss of generality. Also we require Eq.~\eqref{124_positivity1} as necessary conditions. Then, we may rewrite the bound~\eqref{positivity1} as
\begin{align}
&(u_1v_1\sqrt{a_1}-u_2|v_2|\sqrt{a_2})^2
+
\frac{1}{4}(u_1|v_2|-u_2v_1)^2a_4
\nonumber
\\
&
+u_1v_1u_2|v_2|\left[2\sqrt{a_1a_2}+\frac{a_4}{2}+ {\rm sgn}\,(v_2)\left(a_3+\frac{a_4}{2}\right)\right]
\geq0\,,
\end{align}
where sgn denotes the sign function. The inequality is satisfied for arbitrary parameter sets if and only if
\begin{align}
\left|a_3+\frac{a_4}{2}\right|\leq 2\sqrt{a_1a_2}+\frac{a_4}{2}
\,\,\leftrightarrow\,\,
-a_4-2\sqrt{a_1a_2}\leq a_3\leq 2\sqrt{a_1a_2}\,.
\end{align}
Together with Eq.~\eqref{124_positivity1}, this provides the bounds~\eqref{positivity_final}. A map of the allowed region is depicted in the right panel of Fig.~\ref{Fig:envelop}, where we also illustrate that the bound~\eqref{positivity_final} can be obtained by taking envelope of the original bound~\eqref{positivity1}.

\newpage
\bibliography{WGC}{}
\bibliographystyle{utphys}

\end{document}